\documentstyle[11pt,qqa4lart]{article}
\input tcilatex

\QQQ{Language}{
American English
}

\begin{document}

\title{Correlations in interference and diffraction\thanks{%
This project was supported by the National Natural Science Foundation of
China}}
\author{Lu-Ming Duan and Guang-Can Guo\thanks{%
E-mail: gcguo@sunlx06.nsc.ustc.edu.cn} \\
%EndAName
Department of Physics and Nonlinear Science Center, University\\
of Science and Technology of China, Hefei, 230026, P.R.China}
\date{}
\maketitle

\begin{abstract}
\baselineskip  18pt Quantum formalism of Fraunhofer diffraction is obtained.
The state of the diffraction optical field is connected with the state of
the incident optical field by a diffraction factor. Based on this formalism,
correlations of the diffraction modes are calculated with different kinds of
incident optical fields. Influence of correlations of the incident modes on
the diffraction pattern is analyzed and an explanation of the ''ghost''
diffraction is proposed.\\

\end{abstract}

\newpage\baselineskip 18pt

\section{Introduction}

Correlations of states play an important role in quantum cryptography $%
^{\left[ 1,2\right] }$, teleportation $^{\left[ 3,4\right] }$, and
computation $^{\left[ 5-11\right] }$ theory. Correlated states are generated
usually by nonlinear optical processes$^{\left[ 12\right] }$ or by the beam
splitter$^{\left[ 13\right] }$. In this paper, we consider correlations in
interference and diffraction. On the one hand, the diffraction or
interference modes have some interesting correlation properties. On the
other hand, correlations of the incident modes has a notable influence on
the interference or diffraction pattern, in particular, it is the key to the
explanation of the ''ghost'' diffraction$^{\left[ 14\right] }$, an
interesting quantum effect. Interference can be regarded as a special case
of diffraction. To analyze correlations in interference and diffraction, we
need a quantum formalism of diffraction. In the early days of quantum
electrodynamics(QED), it had been proved that the Maxwell equations which
underpin diffraction remain true when the fields are quantized$^{\left[
15-17\right] }$. In quantum optics the entire mode structure of the
diffraction field is still determined by the Helmholtz part of the wave
equation. The role played by quantum mechanics is in determining the states
of the diffraction modes from the states of the incident modes. However, no
systematic approach in determining the states of the diffraction modes has
been proposed. In this paper, we first solve this problem. By introducing
the quantum Kirchhoff boundary condition, we connect the states of the
diffraction modes with the states of the incident modes by a diffraction
factor. Then correlations of the diffraction modes with different kinds of
incident optical fields are calculated. Influence of correlations of the
incident modes on the diffraction pattern is analyzed. The ''ghost'
diffraction is also explained based on this formalism.

We consider Fraunhofer diffraction. This kind of diffraction is most
important. In Section 2, we introduce the equivalent scalar optical field
and the quantum Kirchhoff boundary condition. The equivalent scalar optical
field simplifies the problem of scalar diffraction, in which the variation
of polarization through diffraction is not considered. The quantum Kirchhoff
boundary condition is equivalent in physics to the Kirchhoff boundary
condition in classical scalar diffraction yet overcomes the difficulty that
the classical Kirchhoff boundary condition destroys the commutation
relations of the field operators. In section 3, we obtain quantum formalism
of Fraunhofer diffraction. The normal characteristic functions of the
diffraction modes are connected with those of the incident modes by a
diffraction factor. From the characteristic functions, correlation
properties of the diffraction modes are analyzed in Sec. 4. In this section
the diffraction pattern is also calculated with entangled incident states.
An explanation of the ''ghost'' diffraction is proposed.

\section{The equivalent optical field and the quantum Kirchhoff boundary
condition}

In the diffraction problem the incident and diffraction optical fields are
free. The free quantized electromagnetic field can be expanded into plane
wave modes: 
\begin{equation}
\label{1}\overrightarrow{E}=\stackunder{\overrightarrow{k}}{\sum }%
\stackunder{\mu =1,2}{\sum }i\sqrt{\frac{\hbar \omega }{2V}}a_{%
\overrightarrow{k}\mu }\overrightarrow{e}_{\overrightarrow{k}\mu }e^{i\left( 
\overrightarrow{k}\cdot \overrightarrow{r}-\omega t\right) }+h.c., 
\end{equation}
where $\mu $ is polarization index and $\overrightarrow{k}\cdot $ $%
\overrightarrow{e}_{\overrightarrow{k}\mu }=0$. The annihilation and
creation operators $a_{\overrightarrow{k}\mu },a_{\overrightarrow{k}%
^{^{\prime }}\mu ^{^{\prime }}}^{+}$ satisfy the commutation relation 
\begin{equation}
\label{2}\left[ a_{\overrightarrow{k}\mu },a_{\overrightarrow{k}^{^{\prime
}}\mu ^{^{\prime }}}^{+}\right] =\delta _{\overrightarrow{k}\overrightarrow{k%
}^{^{\prime }}}\delta _{\mu \mu ^{^{\prime }}}. 
\end{equation}

The frequency of the optical field remains unchanged through diffraction. So
we only need consider fields with a definite frequency $\omega $. That is,
in the expansion (1) only the terms with $\left| \overrightarrow{k}\right|
=\frac \omega c$ need be considered. Let $\overrightarrow{k}=\left( k_x,k_y,%
\sqrt{\frac{\omega ^2}{c^2}-k_x^2-k_y^2}\right) $ and $\overrightarrow{k}$
has only two degrees of freedom $k_x,k_y$ (The symbols $\overrightarrow{k}$
below all have this meaning). The incident and diffraction optical fields
are in a half space. Suppose the plane $z=0$ is the diffraction plane, then
there may exist evanescent waves with a depression factor $e^{-\left|
k_z\right| z}$ at both sides of the diffraction plane. So the value domains
of $k_x,k_y$ are $\left( -\infty ,+\infty \right) $, i.e., $k_z$ can be
imaginary. This is different from the plane wave expansions in the whole
space.

In scalar diffraction theory, the boundary condition at the diffraction
plane is independent of the orientation of the optical field, and the
variation of polarization of the optical field through diffraction need not
be considered. So we can introduce the following equivalent scalar optical
field by neglecting the polarization index. 
\begin{equation}
\label{3}\varepsilon \left( \overrightarrow{r}\right) =\frac 1{\sqrt{S}}%
\stackunder{k_x,k_y}{\sum }a_{\overrightarrow{k}}e^{i\overrightarrow{k}\cdot 
\overrightarrow{r}}, 
\end{equation}
where the box-normality of space has been used and $S$ is the cross-section
area of the box. The commutator (2) yields the following commutation
relation of the equivalent optical field at the diffraction plane $z=0$%
\begin{equation}
\label{4}\left[ \varepsilon \left( x,y,0\right) ,\varepsilon ^{+}\left(
x^{^{\prime }},y^{^{\prime }},0\right) \right] =\delta \left( x-x^{^{\prime
}}\right) \delta \left( y-y^{^{\prime }}\right) . 
\end{equation}
In scalar diffraction the equivalent scalar optical field can be in place of
the real optical field. The diffraction problem is much simplified by
introducing the equivalent scalar optical field.

In classical scalar diffraction theory the Kirchhoff boundary condition
states: the optical field remains unchanged through the diffraction aperture 
$\Sigma $ and decays to zero through the diffraction screen$^{[18]}$. This
boundary condition can not be used directly in the quantum case because the
postulate that the optical field decays to zero through the diffraction
screen destroys the commutation relations of the field operators. To keep
consistent with quantum theory, we introduce the following quantum Kirchhoff
boundary condition. The equivalent optical field $\varepsilon \left(
x,y,z=0\right) $ before diffraction is generally in a complicated entangled
state and we use $\rho \left( z=0^{-}\right) $ to represent its whole
density operator. The quantum Kirchhoff boundary condition says: When
passing the diffraction screen all modes of the field $\varepsilon \left(
x,y,z=0\right) $ $\left( x,y\in S-\Sigma \right) $ at the screen undergo
such a strong dissipation that after the screen they are all in the vacuum
state. At the same time, the modes of the field $\varepsilon \left(
x,y,z=0\right) $ $\left( x,y\in \Sigma \right) $ at the aperture undergo no
dissipation at all. From quantum dissipation theory$^{\left[ 19,20\right] }$%
, the total density operator $\rho \left( z=0^{+}\right) $ after diffraction
is expressed as 
\begin{equation}
\label{5}\rho \left( z=0^{+}\right) =tr_{S-\Sigma }\rho \left(
z=0^{-}\right) \otimes \stackunder{\left( x,y\right) \in S-\Sigma }{\prod }%
\left| 0\right\rangle _{xy\text{ }xy}\left\langle 0\right| , 
\end{equation}
where the notation $tr_{S-\Sigma }$ indicates trace of all modes at the
screen. This boundary condition for scalar diffraction is equivalent in
physics to the classical Kirchhoff boundary condition. Yet it is consistent
with quantum mechanics as it results from the quantum dissipation theory. In
next section we use this boundary condition to derive quantum formalism of
diffraction.

\section{Quantum formalism of Fraunhofer diffraction}

In Fraunhofer diffraction the incident and diffraction optical fields are
expanded into the plane wave modes and the role played by quantum mechanics
is in determining the states of the diffraction modes from the states of the
incident modes. Let $a_{\overrightarrow{k}^{^{\prime }}}$ and $b_{%
\overrightarrow{k}}$ represent the annihilation operators of the incident
mode $\overrightarrow{k}^{^{\prime }}$ and the diffraction mode $%
\overrightarrow{k},$ respectively. $\rho \left( a_{\overrightarrow{k}%
_0^{^{\prime }}}\right) $ is the density operator of the incident mode $%
\overrightarrow{k}_0^{^{\prime }}$ and other incident modes are supposed in
the vacuum state. First we derive the reduced normal characteristic function 
$\chi ^{\left( n\right) }\left( b_{\overrightarrow{k}};\xi \right) $ of the
diffraction mode $b_{\overrightarrow{k}}$. Using Eq. (5) and the inverse
transformation of Eq.(3) 
\begin{equation}
\label{6}a_{\overrightarrow{k}}=\frac 1{\sqrt{S}}\int_Sdxdy\varepsilon
\left( \overrightarrow{r}\right) e^{-i\overrightarrow{k}\cdot 
\overrightarrow{r}},
\end{equation}
we get 
\begin{equation}
\label{7}
\begin{array}{c}
\chi ^{\left( n\right) }\left( b_{
\overrightarrow{k}};\xi \right) =\left\langle e^{i\xi ^{*}b_{\overrightarrow{%
k}}^{+}}\cdot e^{i\xi b_{\overrightarrow{k}}}\right\rangle  \\  \\ 
=Tr\left\{ tr_{S-\Sigma }\rho \left( z=0^{-}\right) \otimes 
\stackunder{\left( x,y\right) \in S-\Sigma }{\prod }\left| 0\right\rangle
_{xy\text{ }xy}\left\langle 0\right| \right.  \\ \left. \cdot \exp \left[
i\xi ^{*}\frac 1{
\sqrt{S}}\int_Sdxdy\varepsilon ^{+}\left( x,y,0\right) e^{i\left(
k_xx+k_yy\right) }\right] \cdot \exp \left[ i\xi \frac 1{\sqrt{S}%
}\int_Sdxdy\varepsilon \left( x,y,0\right) e^{-i\left( k_xx+k_yy\right)
}\right] \right\}  \\  \\ 
=Tr\left\{ \rho \left( a_{
\overrightarrow{k}_0^{^{\prime }}}\right) \otimes \stackunder{%
\overrightarrow{k}^{^{\prime }}\neq \overrightarrow{k}_0^{^{\prime }}}{\prod 
}\left| 0\right\rangle _{\overrightarrow{k}^{^{\prime }}\text{ }%
\overrightarrow{k}^{^{\prime }}}\left\langle 0\right| \exp \left\{ i\xi ^{*}%
\stackunder{\overrightarrow{k}^{^{\prime }}}{\sum }a_{\overrightarrow{k}%
^{^{\prime }}}^{+}\frac 1S\int_\Sigma dxdye^{i\left[ \left(
k_x-k_x^{^{\prime }}\right) x+\left( k_y-k_y^{^{\prime }}\right) y\right]
}\right\} \right.  \\ \left. \cdot \exp \left\{ i\xi \stackunder{%
\overrightarrow{k}^{^{\prime }}}{\sum }a_{\overrightarrow{k}^{^{\prime
}}}\frac 1S\int_\Sigma dxdye^{-i\left[ \left( k_x-k_x^{^{\prime }}\right)
x+\left( k_y-k_y^{^{\prime }}\right) y\right] }\right\} \right\} ,
\end{array}
\end{equation}
where $\Sigma $ and $S$ represent area of the diffraction aperture and the
whole diffraction plane, respectively, and the notation $Tr$ indicates trace
of all modes. We define the energy transmissivity $\lambda $ as $\lambda
=\frac \Sigma S.$ Its physical meaning is the ratio of the energy of the
diffraction optical field to the energy of the incident optical field. The
Fraunhofer diffraction factor $f\left( \overrightarrow{k}\right) $ is
defined as 
\begin{equation}
\label{8}f\left( \overrightarrow{k}\right) =\frac{\sqrt{\lambda }}\Sigma
\int_\Sigma e^{-i\left( k_xx+k_yy\right) }dxdy.
\end{equation}
$f\left( \overrightarrow{k}\right) $is normalized by 
\begin{equation}
\label{9}\stackunder{\overrightarrow{k}}{\sum }f^{*}\left( \overrightarrow{k}%
\right) f\left( \overrightarrow{k}\right) =1.
\end{equation}
Eq.(7) is therefore simplified to 
\begin{equation}
\label{10}\chi ^{\left( n\right) }\left( b_{\overrightarrow{k}};\xi \right)
=\chi ^{\left( n\right) }\left[ a_{\overrightarrow{k}_0^{^{\prime }}};\sqrt{%
\lambda }\xi f\left( \overrightarrow{k}-\overrightarrow{k}_0^{^{\prime
}}\right) \right] .
\end{equation}
Eq.(10) connects the reduced normal characteristic function of the
diffraction mode $b_{\overrightarrow{k}}$ with that of the incident mode $a_{%
\overrightarrow{k}_0^{^{\prime }}}$ by a simple diffraction factor.

Similar to the derivation of Eq.(10), the total normal characteristic
function of all diffraction modes $\left\{ b_{\overrightarrow{k}}\right\} $
has the form 
\begin{equation}
\label{11}
\begin{array}{c}
\chi _T^{\left( n\right) }\left( \left\{ b_{
\overrightarrow{k}}\right\} ,\left\{ \xi _{\overrightarrow{k}}\right\}
\right) =\left\langle e^{i\stackunder{\overrightarrow{k}}{\sum }\xi _{%
\overrightarrow{k}}^{*}b_{\overrightarrow{k}}^{+}}e^{i\stackunder{%
\overrightarrow{k}}{\sum }\xi _{\overrightarrow{k}}b_{\overrightarrow{k}%
}}\right\rangle \\  \\ 
=\chi ^{\left( n\right) }\left[ a_{\overrightarrow{k}_0^{^{\prime }}};\sqrt{%
\lambda }\stackunder{\overrightarrow{k}}{\sum }\xi _{\overrightarrow{k}%
}f\left( \overrightarrow{k}-\overrightarrow{k}_0^{^{\prime }}\right) \right] 
\text{ .} 
\end{array}
\end{equation}
The above results are obtained with the supposition that only the incident
mode $\overrightarrow{k}_0^{^{\prime }}$ is not in the vacuum state. If all
the incident modes are in an entangled state, and we use $\chi _T^{\left(
n\right) }\left( \left\{ a_{\overrightarrow{k}^{^{\prime }}}\right\}
,\left\{ \xi _{\overrightarrow{k}^{^{\prime }}}\right\} \right) $ to
indicate its whole normal characteristic function. Eq. (11) can thus be
generalized to 
\begin{equation}
\label{12}\chi _T^{\left( n\right) }\left( \left\{ b_{\overrightarrow{k}%
}\right\} ,\left\{ \xi _{\overrightarrow{k}}\right\} \right) =\chi
_T^{\left( n\right) }\left[ \left\{ a_{\overrightarrow{k}^{^{\prime
}}}\right\} ;\left\{ \sqrt{\lambda }\stackunder{\overrightarrow{k}}{\sum }%
\xi _{\overrightarrow{k}}f\left( \overrightarrow{k}-\overrightarrow{k}%
^{^{\prime }}\right) \right\} \right] . 
\end{equation}
Eq. (12) determines the states of all diffraction modes from the states of
the incident modes. It is a fundamental equation in the quantum formalism of
Fraunhofer diffraction.

The final result (12) is similar to the quantum description of the beam
splitter. For the beam splitter, the input and output modes are linked by a
canonical transformation$^{\left[ 21\right] }$%
\begin{equation}
\label{13}\left( 
\begin{array}{c}
b_1 \\ 
b_2 
\end{array}
\right) =\left( 
\begin{array}{cc}
r & t \\ 
-t & r 
\end{array}
\right) \left( 
\begin{array}{c}
a_1 \\ 
a_2 
\end{array}
\right) , 
\end{equation}
where $a_1,a_2$ are input operators and $b_1,b_2$ are output operators. The
parameters $r$ and $t$ should satisfy $r^2+t^2=1.$ From Eq. (13), we obtain
the relation of the normal characteristic function between the input and
output modes 
\begin{equation}
\label{14}\chi ^{\left( n\right) }\left( b_1,b_2;\xi _1,\xi _2\right) =\chi
^{\left( n\right) }\left( a_1,a_2;r\xi _1-t\xi _2,t\xi _1+r\xi _2\right) . 
\end{equation}
Eqs. (14) and (12) are very alike in the form. However, some important
differences lie in their derivation. In diffraction the input and output
modes cannot be put in a canonical transformation, which may be seen from
the relation 
\begin{equation}
\label{15}\stackunder{\overrightarrow{k}}{\sum }\sqrt{\lambda }f^{*}\left( 
\overrightarrow{k}-\overrightarrow{k}^{^{\prime }}\right) \sqrt{\lambda }%
f\left( \overrightarrow{k}-\overrightarrow{k}^{^{\prime }}\right) =\lambda <1%
\text{ .} 
\end{equation}
Only when the energy transmissivity $\lambda =\frac \Sigma S=1$ , i.e., when
there is no diffraction screen, the input and output modes can be linked by
a trivial canonical transformation. So unlike Eq. (14), Eq. (12) is not a
direct result of the input-output theory$^{\left[ 22,19\right] }$. In the
derivation of Eq. (12), the quantum Kirchhoff boundary condition plays an
essential role.

The general equation (12) can describe interference as well as diffraction.
If there are two diffraction apertures $\Sigma _1,\Sigma _2$ , the
diffraction factor $f\left( \overrightarrow{k}\right) $ simply becomes 
\begin{equation}
\label{16}f\left( \overrightarrow{k}\right) =\frac{\sqrt{\lambda }}{\Sigma
_1+\Sigma _2}\int_{\Sigma _1+\Sigma _2}e^{-i\left( k_xx+k_yy\right) }dxdy, 
\end{equation}
where $\lambda =\frac{\Sigma _1+\Sigma _2}S$ . When $\Sigma _1,\Sigma _2$
tend to zero, Eq. (12) with this $f\left( \overrightarrow{k}\right) $ gives
quantum description of the double-slit interference.

\section{Correlations in interference and diffraction}

\subsection{correlations of the diffraction (or interference) modes}

In this subsection we consider correlations of the diffraction modes.
Suppose all the incident modes except $\overrightarrow{k}_0^{^{\prime }}$
are in the vacuum state. First we show that the diffraction modes are not
correlated only when the incident mode $\overrightarrow{k}_0^{^{\prime }}$
is in a coherent state. If the diffraction modes are independent, the
decomposition $\chi _T^{\left( n\right) }\left( \left\{ b_{\overrightarrow{k}%
}\right\} ,\left\{ \xi _{\overrightarrow{k}}\right\} \right) =\stackunder{%
\overrightarrow{k}}{\prod }\chi ^{\left( n\right) }\left( b_{\overrightarrow{%
k}},\xi _{\overrightarrow{k}}\right) $ should hold. From Eq. (11) this
decomposition holds if and only if $\chi ^{\left( n\right) }\left[ a_{%
\overrightarrow{k}_0^{^{\prime }}};\xi \right] $ has the following form 
\begin{equation}
\label{17}\chi ^{\left( n\right) }\left[ a_{\overrightarrow{k}_0^{^{\prime
}}};\xi \right] =e^{i\left( \xi ^{*}\alpha ^{*}+\xi \alpha \right) }, 
\end{equation}
i.e., the incident mode is in a coherent state. Under this condition, the
diffraction modes are not correlated and all in coherent states. With any
other kinds of incident optical fields the diffraction modes are in an
entangled state.

The above discussion shows that the diffraction modes are generally
correlated. In experiments correlation of the photon number is widely used,
so we first calculate the correlation coefficient of the photon number of
two diffraction modes .The correlation coefficient is defined by 
\begin{equation}
\label{18}\eta =\frac{\left\langle \Delta n_{\overrightarrow{k}_1}\Delta n_{%
\overrightarrow{k}_2}\right\rangle }{\left[ \left\langle \left( \Delta n_{%
\overrightarrow{k}_1}\right) ^2\right\rangle \left\langle \left( \Delta n_{%
\overrightarrow{k}_2}\right) ^2\right\rangle \right] ^{\frac 12}}, 
\end{equation}
where $n_{\overrightarrow{k}_i}$ $\left( i=1,2\right) $ denotes the number
operator of the mode $\overrightarrow{k}_i$. After some calculation, from
Eq. (11) we obtain 
\begin{equation}
\label{19}\eta =\frac{F_n-1}{\left[ \left( F_n+h_1-1\right) \left(
F_n+h_2-1\right) \right] ^{\frac 12}}, 
\end{equation}
where $F_n$ is the Fano factor of the incident mode $\overrightarrow{k}%
_0^{^{\prime }}$ , i.e., 
\begin{equation}
\label{20}F_n=\frac{\left\langle \left( \Delta n_{\overrightarrow{k}%
_0^{^{\prime }}}\right) ^2\right\rangle }{\left\langle n_{\overrightarrow{k}%
_0^{^{\prime }}}\right\rangle }, 
\end{equation}
and $h_i$ in Eq. (19) is defined by 
\begin{equation}
\label{21}h_i=\frac 1{\lambda \left| f\left( \overrightarrow{k}_i-%
\overrightarrow{k}_0^{^{\prime }}\right) \right| ^2}\text{ }\left(
i=1,2\right) . 
\end{equation}
The relation between $\eta $ and $F_n$ is illustrated in Fig. 1\\

\begin{center}
Fig. 1\\
\end{center}

If the incident mode is in a thermal state, $F_n=$ $\left\langle n_{%
\overrightarrow{k}_0^{^{\prime }}}\right\rangle +1$ and $\eta $ tends to its
maximum value $1$ with $\left\langle n_{\overrightarrow{k}_0^{^{\prime
}}}\right\rangle >>1$ . The correlation coefficient $\eta $ gets its minimum
value $-\left[ \left( h_1-1\right) \left( h_2-1\right) \right] ^{-\frac 12}$
with the incident mode in a Fock state.

Though $\eta \approx 1$ if the incident mode is in a thermal state with $%
\left\langle n_{\overrightarrow{k}_0^{^{\prime }}}\right\rangle >>1$, the
diffraction modes are not correlated perfectly in this case. That can be
seen from residual variance of the variables $n_{\overrightarrow{k}_1}$ and $%
n_{\overrightarrow{k}_2}$ in the linear regression. The residual variance of
the variable $n_{\overrightarrow{k}_1}$ has the form$^{\left[ 23\right] }$%
\begin{equation}
\label{22}
\begin{array}{c}
Var\left( n_{
\overrightarrow{k}_1}-\beta _1n_{\overrightarrow{k}_2}-\beta _2\right)
=\left\langle \left( \Delta n_{\overrightarrow{k}_1}\right) ^2\right\rangle
\left( 1-\eta ^2\right) \\ =\frac{\left\langle n_{\overrightarrow{k}%
_0^{^{\prime }}}\right\rangle }{h_1^2}\left( F_n+h_1-1\right) \left( 1-\eta
^2\right) , 
\end{array}
\end{equation}
where $\beta _1$ and $\beta _2$ are linear regression coefficients. Suppose $%
h_1=h_2$ and $\left\langle n_{\overrightarrow{k}_0^{^{\prime
}}}\right\rangle >>1,$ then 
\begin{equation}
\label{23}Var\left( n_{\overrightarrow{k}_1}-\beta _1n_{\overrightarrow{k}%
_2}-\beta _2\right) \approx \frac{2\left\langle n_{\overrightarrow{k}%
_0^{^{\prime }}}\right\rangle }{h_1}=2\left\langle n_{\overrightarrow{k}%
_1}\right\rangle . 
\end{equation}
So in this case the residual variance is very large. In fact, the equation $%
\eta \rightarrow 1$ under the condition $\left\langle n_{\overrightarrow{k}%
_0^{^{\prime }}}\right\rangle \rightarrow \infty $ results from the infinite
variance of $n_{\overrightarrow{k}_1}$ and $n_{\overrightarrow{k}_2}$ .We
can not conclude from $\eta \rightarrow 1$ that the diffraction modes are
correlated perfectly.

For the beam splitter, the correlation of the number operator of the output
modes has the same form as Eq.(19). However, there are still some
differences. First, the equation $\frac 1{h_1}+$ $\frac 1{h_2}=1$ holds for
the beam splitter whereas in diffraction we have $\frac 1{h_1}+$ $\frac
1{h_2}<1$ . So for the beam splitter, the correlation coefficient of the
output number operators can attain its minimum value -1 with the input mode
in a Fock state. Second, in diffraction or interference correlations of many
modes can be generated whereas the beam splitter is only used to prepare
two-mode entangled states.

Correlation coefficients describe correlation properties of a pair of
specialized operators. Several approaches to the description of quantum
entanglement have been proposed. In particular, Schlienz and Mahler
interpreted the difference between the entangled state and the product state
as the entanglement$^{\left[ 24\right] }$. Suppose $\rho $ is the density
operator of the whole system and $\rho _a=tr_b\rho $ , $\rho _b=tr_a\rho $ ,
where the subscripts $a$ and $b$ represent two subsystems. The
Schlienz-Mahler measure is defined by$^{\left[ 24\right] }$ 
\begin{equation}
\label{24}\gamma =\sqrt{\frac{N\left( \rho \right) }{N\left( \rho \right) -1}%
tr\left[ \left( \rho -\rho _a\otimes \rho _b\right) ^2\right] ,} 
\end{equation}
where $N\left( \rho \right) $ indicates the dimension of the density
operator $\rho $ and $\gamma $ defined above satisfies $0\leq \gamma \leq 1.$
However, the more recent papers distinguish quantum entanglement from
classical correlations$^{\left[ 25-30\right] }$. The entanglement is
interpreted as the degree of inseparability. The entangled state is said to
be inseparable if it can not be expressed as a mixture of product states of
two subsystems. In this interpretation, the Schlienz-Mahler quantity $\gamma 
$ measures the total correlations rather than pure quantum entanglement. It
is now believed that pure quantum entanglement can not be fully described by
a single quantity$^{\left[ 29\right] }$. Bennett et. al. defined two
quantities:$^{\left[ 26,29\right] }$ ''entanglement of formation'' defined
as the least number of shared singlets asymptotically required to prepare $%
\rho $ by local operations and classical communication, and ''distillable
entanglement'' defined as the greatest number of pure singlets that can
asymptotically be prepared from $\rho $ by local operations and classical
communication. And recently, Vedral et. al. introduced a new measure of
entanglement$^{\left[ 30\right] }$, which interprets the entanglement as the
minimum distance to all separable states. These measures have the desirable
feature that their expectations can not be increased by local operations,
but the disadvantage of being hard to evaluate because of the implied
optimization. The question is still open in this direction.

Though the Schlienz-Mahler quantity $\gamma $ in fact measures the total
correlations, it is superior to the correlation coefficients, since it is
not limited to specialized observables. In the following we use the
Schlienz-Mahler quantity to analyze correlation properties of the
diffraction modes. Before doing this, we first introduce the following lemma.

{\bf Lemma. }Suppose $\rho _{1,}\rho _2$ are two density operators of boson
fields, and $\chi _1^{\left( n\right) }\left( \xi \right) ,$ $\chi
_2^{\left( n\right) }\left( \xi \right) $ are normal characteristic
functions of $\rho _1$ and $\rho _2,$ respectively, then we have 
\begin{equation}
\label{25}tr\left( \rho _1\rho _2\right) =\int \frac{d^2\xi }\pi \chi
_1^{\left( n\right) }\left( \xi \right) \chi _2^{\left( n\right) }\left(
-\xi \right) e^{-\left| \xi \right| ^2}. 
\end{equation}

{\bf Proof.} If generalized functions (such as derivatives of delta
functions) are permitted, the existence proof of P-functions of Boson fields
has been given by Klauder and Sudarshan$^{\left[ 31,19\right] }$. So $%
tr\left( \rho _1\rho _2\right) $ can be expressed as 
\begin{equation}
\label{26}
\begin{array}{c}
tr\left( \rho _1\rho _2\right) =\int P_1\left( \alpha \right) \left\langle
\alpha \right| \rho _2\left| \alpha \right\rangle d^2\alpha \\  
\\ 
=\pi \int P_1\left( \alpha \right) Q_2\left( \alpha \right) d^2\alpha , 
\end{array}
\end{equation}
where $P_1\left( \alpha \right) $ and $Q_2\left( \alpha \right) $ are
P,Q-functions of the density operators $\rho _{1,}\rho _2$ respectively. The
P,Q-functions are Fourier transformations of the normal and anti-normal
characteristic functions. So Eq. (26) can be rewritten as 
\begin{equation}
\label{27}tr\left( \rho _1\rho _2\right) =\int \frac{d^2\xi }\pi \chi
_1^{\left( n\right) }\left( \xi \right) \chi _2^{\left( a\right) }\left(
-\xi \right) , 
\end{equation}
where $\chi ^{\left( a\right) }\left( \xi \right) $ indicates the
anti-normal characteristic function. Eq. (27) is equivalent to Eq. (25) .
This completes the proof.

We calculate the Schlienz-Mahler quantity $\gamma $ with a thermal incident
optical field. From Eq. (11) the normal characteristic function of the
diffraction modes $\overrightarrow{k}_1$ and $\overrightarrow{k}_2$ has the
form 
\begin{equation}
\label{28}\chi _T^{\left( n\right) }\left( \left\{ b_{\overrightarrow{k}%
_1},b_{\overrightarrow{k}_2}\right\} ;\left\{ \xi _{\overrightarrow{k}%
_1},\xi _{\overrightarrow{k}_2}\right\} \right) =e^{-\left\langle
N\right\rangle \sqrt{\lambda }\left[ \xi _{\overrightarrow{k}_1}f\left( 
\overrightarrow{k}_1-\overrightarrow{k}_0^{^{\prime }}\right) +\xi _{%
\overrightarrow{k}_2}f\left( \overrightarrow{k}_2-\overrightarrow{k}%
_0^{^{\prime }}\right) \right] }, 
\end{equation}
where $\left\langle N\right\rangle $ is the mean photon number of the
incident mode. For thermal states, the dimension of the density operator $%
N\left( \rho \right) \rightarrow \infty $. Eq. (24) together with Eq. (25)
yields 
\begin{equation}
\label{29}\gamma ^2=\frac 1{x_1x_2-4y^2}-\frac 2{x_1x_2-y^2}+\frac
1{x_1x_2}, 
\end{equation}
where 
\begin{equation}
\label{30}x_i=2\left\langle N\right\rangle \lambda \left| f\left( 
\overrightarrow{k}_i-\overrightarrow{k}_0^{^{\prime }}\right) \right| ^2+1%
\text{ }\left( i=1,2\right) , 
\end{equation}
\begin{equation}
\label{31}y=\left\langle N\right\rangle \lambda \left| f\left( 
\overrightarrow{k}_1-\overrightarrow{k}_0^{^{\prime }}\right) f\left( 
\overrightarrow{k}_2-\overrightarrow{k}_0^{^{\prime }}\right) \right| 
\end{equation}
From Eq. (29) it is obvious that $\gamma $ tends to zero if $\left\langle
N\right\rangle \rightarrow \infty $ or $\left\langle N\right\rangle
\rightarrow 0.$ If the diffraction factor satisfies $\left| f\left( 
\overrightarrow{k}_1-\overrightarrow{k}_0^{^{\prime }}\right) \right|
=\left| f\left( \overrightarrow{k}_2-\overrightarrow{k}_0^{^{\prime
}}\right) \right| $ , we have $x_1=x_2=2y+1$ and $\gamma $ is simplified to 
\begin{equation}
\label{32}\gamma =\left[ \frac 1{4y+1}-\frac 2{\left( 3y+1\right) \left(
y+1\right) }+\frac 1{\left( 2y+1\right) ^2}\right] ^{\frac 12}. 
\end{equation}
The relation between $\gamma $ and $y$ is illustrated in Fig. 2\\

\begin{center}
Fig. 2\\
\end{center}

From the figure we see the Schlienz-Mahler quantity $\gamma $ attains the
maximum when $y\approx 1.1$. The maximum value is $0.25$. With a larger or
smaller mean photon number, the correlation of the diffraction modes
decreases.

\subsection{Influence of correlations of the incident modes on the
diffraction (or interference) pattern}

To show the influence of correlations on the diffraction pattern, we
consider the circumstance with two incident modes being in an entangled
state. The entangled state is prepared by a beam splitter with the input
mode in a Fock state. From Eq. (14) the normal characteristic function of
the two incident modes has the form 
\begin{equation}
\label{33}\chi _T^{\left( n\right) }\left( a_{\overrightarrow{k}_1^{^{\prime
}}},a_{\overrightarrow{k}_2^{^{\prime }}};\xi _{\overrightarrow{k}%
_1^{^{\prime }}},\xi _{\overrightarrow{k}_2^{^{\prime }}}\right) =e^{i\left(
r^{*}\xi _{\overrightarrow{k}_1^{^{\prime }}}^{*}-t^{*}\xi _{\overrightarrow{%
k}_2^{^{\prime }}}^{*}\right) a^{+}}e^{i\left( r\xi _{\overrightarrow{k}%
_1^{^{\prime }}}-t\xi _{\overrightarrow{k}_2^{^{\prime }}}\right) a}\left|
n\right\rangle , 
\end{equation}
where $a$ denotes the input mode of the beam splitter. The diffraction
pattern is shown by the mean photon number distribution $\left\langle n_{%
\overrightarrow{k}}\right\rangle $ of the diffraction modes. Eq. (12) gives 
\begin{equation}
\label{34}
\begin{array}{c}
\left\langle n_{
\overrightarrow{k}}\right\rangle =\lambda \left\{ \left| f\left( 
\overrightarrow{k}-\overrightarrow{k}_1^{^{\prime }}\right) \right|
^2\left\langle n_{\overrightarrow{k}_1^{^{\prime }}}\right\rangle +\left|
f\left( \overrightarrow{k}-\overrightarrow{k}_2^{^{\prime }}\right) \right|
^2\left\langle n_{\overrightarrow{k}_2^{^{\prime }}}\right\rangle \right. \\
\\ 
\left. +\left[ f^{*}\left( 
\overrightarrow{k}-\overrightarrow{k}_1^{^{\prime }}\right) f\left( 
\overrightarrow{k}-\overrightarrow{k}_2^{^{\prime }}\right) \left\langle a_{%
\overrightarrow{k}_1^{^{\prime }}}^{+}a_{\overrightarrow{k}_2^{^{\prime
}}}\right\rangle +h.c.\right] \right\} \\  \\ 
=\lambda n\left| rf\left( \overrightarrow{k}-\overrightarrow{k}_1^{^{\prime
}}\right) -tf\left( \overrightarrow{k}-\overrightarrow{k}_2^{^{\prime
}}\right) \right| ^2. 
\end{array}
\end{equation}
If the two incident modes are not correlated, i.e., if they are represented
by the density operator $tr_{\overrightarrow{k}_1^{^{\prime }}}\rho \otimes
tr_{\overrightarrow{k}_2^{^{\prime }}}\rho $, the mean photon number
distribution of the diffraction modes becomes 
\begin{equation}
\label{35}\left\langle n_{\overrightarrow{k}}\right\rangle =\lambda n\left[
\left| rf\left( \overrightarrow{k}-\overrightarrow{k}_1^{^{\prime }}\right)
\right| ^2+\left| tf\left( \overrightarrow{k}-\overrightarrow{k}_2^{^{\prime
}}\right) \right| ^2\right] . 
\end{equation}
So the two incident modes are superposed coherently if they are correlated
and incoherently if they are not.

The influence of correlations on the diffraction (or interference) is
dramatically illustrated by the ''ghost'' diffraction (or interference)
effect. In the observation experiment of the ''ghost'' diffraction$^{\left[
14\right] }$, a light beam, which is generated from spontaneous parametric
down-conversion (SPDC) and consists of two orthogonal polarization
components (usually called signal and idler), is split by a polarization
beam splitter into two beams, and detected by two distinct pointlike
photoncounting detectors for coincidence. A Young's double-slit or
single-slit aperture is inserted into the {\it signal} beam. Surprisingly,
an interference or diffraction pattern is observed in the coincidence counts
by scanning the detector in the {\it idler} beam. Here we give an exact
explanation of the ''ghost'' diffraction. For the SPDC, the output light is
in a superposition of the vacuum and two-photon states$^{\left[ 12\right] }$%
\begin{equation}
\label{36}\left| \Psi \right\rangle =\left| 0\right\rangle +F\stackunder{%
\overrightarrow{k}^{^{\prime }}}{\sum }a_{\overrightarrow{k}^{^{\prime
}}}^{+}c_{\overrightarrow{k}^{^{\prime }}}^{+}\left| 0\right\rangle , 
\end{equation}
where the operators $a_{\overrightarrow{k}^{^{\prime }}}^{+}$ and $c_{%
\overrightarrow{k}^{^{\prime }}}^{+}$ represent the signal and idler modes,
respectively. The normal characteristic function of the signal and idler
modes is indicated by $\chi _T^{\left( n\right) }\left( \left\{ a_{%
\overrightarrow{k}^{^{\prime }}}\right\} ,\left\{ c_{\overrightarrow{k}%
^{^{\prime }}}\right\} ;\left\{ \xi _{1\overrightarrow{k}^{^{\prime
}}}\right\} ,\left\{ \xi _{2\overrightarrow{k}^{^{\prime }}}\right\} \right) 
$. Then the signal light meets a diffraction screen and the idler light
remains unchanged. The second-order correlation coefficient between a fixed
diffraction mode and arbitrary idler modes is to be measured. From Eq. (12),
the normal characteristic function of the diffraction and idler modes has
the form 
\begin{equation}
\label{37}
\begin{array}{c}
\chi _T^{\left( n\right) }\left( \left\{ b_{
\overrightarrow{k}}\right\} ,\left\{ c_{\overrightarrow{k}^{^{\prime
}}}\right\} ;\left\{ \xi _{\overrightarrow{k}}\right\} ,\left\{ \xi _{2%
\overrightarrow{k}^{^{\prime }}}\right\} \right) \\  \\ 
=\chi _T^{\left( n\right) }\left( \left\{ a_{\overrightarrow{k}^{^{\prime
}}}\right\} ,\left\{ c_{\overrightarrow{k}^{^{\prime }}}\right\} ;\left\{ 
\sqrt{\lambda }\stackunder{\overrightarrow{k}}{\sum }\xi _{\overrightarrow{k}%
}f\left( \overrightarrow{k}-\overrightarrow{k}^{^{\prime }}\right) \right\}
,\left\{ \xi _{2\overrightarrow{k}^{^{\prime }}}\right\} \right) . 
\end{array}
\end{equation}
With Eqs. (36) and (37), we obtain the second-order correlation coefficient
between a fixed diffraction mode $\overrightarrow{k}$ and an arbitrary idler
mode $\overrightarrow{k}^{^{\prime }}$%
\begin{equation}
\label{38}G^{\left( 2\right) }\left( b_{\overrightarrow{k}},c_{%
\overrightarrow{k}^{^{\prime }}}\right) =\left\langle b_{\overrightarrow{k}%
}^{+}b_{\overrightarrow{k}}c_{\overrightarrow{k}^{^{\prime }}}^{+}c_{%
\overrightarrow{k}^{^{\prime }}}\right\rangle =\lambda \left| F\right|
^2\left| f\left( \overrightarrow{k}-\overrightarrow{k}^{^{\prime }}\right)
\right| ^2. 
\end{equation}
It is directly proportional to square of the diffraction factor. The
diffraction pattern occurs by fixing the diffraction mode and scanning the
idler modes. Therefore, Eq. (38) explains the observation in the ''ghost''
diffraction.

\newpage\ 

{\bf Caption 1:} The relation between the correlation coefficient $\eta $ of
the diffraction modes and the Fano factor $F_n$ of the incident mode. We
choose $h_1=h_2=3$.\\

{\bf Caption 2:} The relation between the Schlienz-Mahler quantity $\gamma $
and the mean photon number of the incident mode. $y$ is expressed by Eq.(31).


\newpage
\baselineskip  20pt

\begin{thebibliography}{99}
\bibitem{1}  A. Ekert, Phys. Rev. Lett. 67, 661 (1991).

\bibitem{2}  C.H. Bennett, Phys. Rev. Lett. 68, 3121 (1992).

\bibitem{3}  C.H. Bennett, G. Brassard, C. Crepeau, et.al., Phys. Rev. Lett.
70, 1895 (1993).

\bibitem{4}  S.L. Braunstein, Phys. Rev. A 53, 1900 (1996).

\bibitem{5}  D. Deutsch and R. Jozsa, Proc. R. Soc. London Ser. A 439, 553
(1992).

\bibitem{6}  A. Barenco, Contemporary Phys. 37, No. 5, 375 (1996).

\bibitem{7}  A. Ekert and R. Jozsa, Rev. Mod. Phys. 68, No. 3, 733 (1996).

\bibitem{8}  J.I. Cirac and P. Zoller, Phys. Rev. Lett. 74, 4091 (1995).

\bibitem{9}  W.G. Unruh, Phys. Rev. A 51, 992 (1995).

\bibitem{10}  G.M. Palma, K.A. Suominen, and A.K. Ekert, Proc. R. Soc.
London A 452, 567 (1996).

\bibitem{11}  M.B. Plenio and P.L. Knight, Phys. Rev. A 53, 2986 (1996).

\bibitem{12}  M.H. Rubin, D.N. Klyshko, Y.H. Shih and A.V. Sergienko, Phys.
Rev. A 50, 5122 (1994).

\bibitem{13}  M. Reck, A. Zeilinger, H.J. Berstein and P. Bertani, Phys.
Rev. Lett. 73, 58 (1994).

\bibitem{14}  D.V. Sterkalov, A.V. Sergienko, D.N. Klyshko, and Y.H. Shih,
Phys. Rev. Lett. 74, 3500 (1995).

\bibitem{15}  G. Racah, Lincei Rend. 11, 837, 1100 (1930).

\bibitem{16}  W. Heisenberg, Ann. d. Physik 9, 338 (1931).

\bibitem{17}  E. Fermi, Rev. Mod. Phys. 4, 87 (1932).

\bibitem{18}  M. Born and E. Wolf, the Principles of Optics, 3rd edn.
Pergamon, London (1965).

\bibitem{19}  C.W. Gardiner, Quantum Noise, Springer Verlag, Berlin
Heidelberg (1991).

\bibitem{20}  L.M. Duan and G.C. Guo, Chin. Phys. Lett. 121, 589 (1995).

\bibitem{21}  H. Huttner and Y. Ben-Aryeh, Phys. Rev. A 38, 204 (1988).

\bibitem{22}  L. Hilico, C. Fabre, S. Reynaud and E. Giacobino, Phys. Rev. A
46, 4396 (1992).

\bibitem{23}  M. Fisz, Probability Theory and Mathematical Statistics, VEB
Deutscher Verlag der Wissenschaften, Berlin (1958).

\bibitem{24}  J. Schlienz, and G. Mahler, Phys. Rev. A 52, 4396 (1995).

\bibitem{25}  S. Popescu, Phys. Rev. Lett. 72, 797 (1994); 74, 2619 (1995).

\bibitem{26}  C.H. Bennett, H.J. Bernstein, S. Popescu and B. Schumacher,
Phye. Rev. A 53, 2046 (1996).

\bibitem{27}  P. Horodecki and R. Horodecki, Phys. Rev. Lett. 76, 2196
(1996).

\bibitem{28}  A. Peres, Phys. Rev. Lett. 77, 1413 (1996).

\bibitem{29}  C. H.Bennett, D.P. DiVincenzo, J.A. Smolin, and W.K. Wootters,
Phys. Rev. A 54, 3824 (1996).

\bibitem{30}  V. Vedral, M.B. Plenio, M.A. Rippin, and P.L. Knight, Phys.
Rev. Lett. 78, 2275 (1997).

\bibitem{25}  J.R. Klauder, E.C.G. Sudarshan, Fundamentals of Quantum
Optics, Benjamin, New York (1968).
\end{thebibliography}
\end{document}